\documentclass[preprintnumbers,amsmath,amssymb,aps,eqsecnum]{revtex4}
\usepackage[dvips]{graphics,color}
\usepackage{latexsym}
\usepackage{graphicx}
\def\be{\begin{equation}}
\def\ee{\end{equation}}
\def\bea{\begin{eqnarray}}
\def\eea{\end{eqnarray}}

\begin{document}
\title{Structure formation in modified gravity models alternative to dark energy} 
\author{Kazuya Koyama} 

\affiliation{\vspace*{0.2cm} Institute of Cosmology \& 
Gravitation, University of Portsmouth, Portsmouth~PO1~2EG, UK 
\vspace*{0.2cm}} 
\begin{abstract}
We study structure formation in phenomenological models 
in which the Friedmann equation receives a correction of the form 
$H^{\alpha}/r_c^{2-\alpha}$, which realize an accelerated expansion without 
dark energy. 
In order to address structure formation
in these model, we construct simple covariant gravitational equations 
which give the modified Friedmann equation with $\alpha=2/n$ where $n$ is an integer. 
For $n=2$, the underlying theory is known as a 5D braneworld model (the DGP model). 
Thus the models interpolate between the DGP model ($n=2, \alpha=1$) and the
LCDM model in general relativity ($n \to \infty, \alpha \to 0$). 
Using the covariant equations, cosmological perturbations are analyzed. 
It is shown that in order to satisfy the Bianchi identity at a perturbative level, 
we need to introduce a correction term $E_{\mu \nu}$ in the effective equations. 
In the DGP model, $E_{\mu \nu}$ comes from 5D gravitational fields and correct conditions on 
$E_{\mu \nu}$ can be derived by solving the 5D perturbations. In the general case $n>2$, 
we have to assume the structure of a modified theory of gravity 
to determine $E_{\mu \nu}$. We show that structure formation is 
different from a dark energy model in general relativity with identical 
expansion history and that quantitative features of the difference 
crucially depend on the conditions on $E_{\mu \nu}$, that is, 
the structure of the underlying theory of modified gravity.
This implies that it is essential to identify underlying theories
in order to test these phenomenological models against observational data
and, once we identify a consistent theory, structure formation tests become 
essential to distinguish modified gravity models from dark energy models in 
general relativity. 

\end{abstract}
\maketitle

\section{Introduction}
The discovery of cosmic acceleration presents a 
deep puzzle for cosmology \cite{SN}. A conventional way to explain this fact is to 
introduce a tiny cosmological constant or dark energy in the context 
of general relativity. However, it is also possible to think that the 
standard Friedmann equation which determines the expansion of the 
universe is modified. There have been many attempts to modify the 
Friedmann equation either empirically or based on a modified 4-dimensional 
action \cite{T} and study observational constraints coming from the expansion history 
of the universe. 

One example of the explicit realization of the modified Friedmann equation
is provided by the Dvali-Gabadadze-Porrati (DGP) brane-world model \cite{Dvali:2000rv},  
in which gravity leaks off the 4-dimensional brane into the 5-dimensional ``bulk"
Minkowski spacetime at large scales. The energy conservation equation remains the same as in
general relativity, but the Friedman equation is modified:
\be
{H \over r_c}= H^2-{8\pi G \over 3}\rho\,. 
\label{Fried}
\ee
The modified Friedman equation~(\ref{Fried}) shows that at late times
we have $H\to H_\infty= 1/r_c$. 
Since $H_0>H_\infty$, in order to achieve acceleration at late
times, we require $r_c\gtrsim H_0^{-1}$, and this is confirmed by
fitting SN observations \cite{Deffayet:2002sp}.
Although it has been shown that the DGP model suffers from serious 
theoretical problems, such as the existence of a ghost in de Sitter 
solutions of Eq.~(\ref{Fried}) and the strong coupling problem \cite{problems, problems2}, 
the DGP model is the simplest covariant theory for  
modified gravity which gives accelerated expansion of the universe without dark energy.
In addition, the DGP model allows us to determine how modified gravity affects 
various cosmological observations other than the modified expansion history of the universe. 
It is important to stress that in the DGP model the modification to the Friedman
equation is derived from a covariant 5-dimensional action and
junction conditions across the brane. Thus it is possible to derive 
4-dimensional covariant effective equations which govern the dynamics of 
gravity on the brane. 

The expansion history of the DGP model is quite different
from the LCDM model. The expansion history of the DGP is equivalent to that in 
dark energy models with an equation of state $w(a)=-1/(1+\Omega_m(a))$ 
where $\Omega_m(a)$ is the density parameter for matter \cite{Lue:2004rj}. 
This connection between $w$ and $\Omega_m$ can be used to impose strong 
constraints on the models. If $\Omega_{m}(a_0) \sim 0.3$, 
the expansion history of the DGP is equivalent to that of dark energy models 
with $w > -1$. Given the fact that the present SNe data favors LCDM or 
even prefers $w < -1$, the expansion history alone may be enough to 
falsify the DGP model. Indeed, Ref.~\cite{Fairbairn} showed that, combining the data from 
Supernova Legacy Survey (SNLS) and the baryon acoustic peak in the Sloan
Digital Sky Survey data, the DGP model is not compatible with 
a spatially flat universe. This is confirmed by Ref.~\cite{Alam} where 
it is shown that the same data sets exclude the flat DGP model at $3 \sigma$. 
On the other hand, Ref.~\cite{Alam} also showed that if we use the 'Gold' data for supernovae, 
the flat DGP model is still acceptable at $2 \sigma$ level. Together with 
the current status of measurements for $\Omega_m$ from galaxy surveys, 
it may be too early to reject the flat DGP model from observations. However, it is 
true that the flat DGP model is in tension with the data purely 
from the background dynamics. 

Then it is tempting to consider models that give different expansion histories from the 
DGP, like the quintessence models in general relativity. 
Inspired by the DGP model, the modified Friedmann equation of the form
\be
{H^{\alpha} \over r_c^{2-\alpha}} = H^2 - \frac{8 \pi G}{3} \rho.
\label{alpha}
\ee
is proposed by Dvali and Turner \cite{DT}. This is the simplest generalization of the idea of the 
self-accelerating universe where there exists a single dimensionfull 
parameters $r_c$. Thus for a given $\alpha$,
the model still has the same number of parameters as LCDM. Ref.~\cite{Fairbairn} 
also reported the constraint on $\alpha$ ($-0.8 < \alpha < 0.3$ at $1 \sigma$)
from SNLS data. 
A problem of these phenomenological approaches is that it is 
difficult to discuss the full observational consequences.
If we modify the Friedmann equation in the background, it is natural 
to think that behaviour of perturbations around this background is
different from that in general relativity. 
However, without specifying self-consistent models that implement the modification 
of the Friedmann equation, it is difficult to study perturbations 
in a consistent way.

The aim of this paper is moderate and we do not try to construct
the underlying models. Instead, we study how the consistency of 
the gravitational equations can constrain the behaviour of perturbations 
around the modified Friedmann equation (\ref{alpha}). For this purpose
it is desirable to derive effective covariant equations which reproduce the modified 
Friedmann equation in the cosmological background. 
We construct the simplest possible covariant gravitational equations 
by extending the covariant effective 
equations in the DGP model derived in Ref.~\cite{Maeda:2003ar} based on 
the method proposed by Ref.~\cite{Shiromizu:1999wj}. 
A proposed covariant equation is written in the form
\be
G_{\mu \nu} = \kappa_{(n)}^4 {}^{(n)}\Pi_{\mu \nu} -E_{\mu \nu},
\ee
where $^{(n)}\Pi_{\mu \nu}$ is a $n$-th power function of  
$\kappa_4^{-2} G_{\mu \nu}-  T_{\mu \nu}$ and 
\be
\kappa_{(n)}^4= 4 r_c^{2(n-1)} \kappa_4^{2n},
\ee
where $\kappa_4^2 = 8 \pi G$.
We assume that the conservation of the energy-momentum tensor holds,
$\nabla^{\mu} T_{\mu \nu}=0$. The $E_{\mu \nu}$ tensor is necessary to 
satisfy the Bianchi identity.

The structure of this paper is as follows. In section II, 
we show that, in the cosmological background, the effective equations 
reproduce the modified Friedmann equation (\ref{Fried}) with $\alpha = 2/n$
where $n$ is an integer. 
In the background spacetime, it is possible to satisfy the 
Bianchi identity without $E_{\mu \nu}$. In section III, it is shown that, 
at a perturbative level, we need perturbations of $E_{\mu \nu}$ to satisfy 
the Bianchi identity.
Without an underlying model, the perturbation of $E_{\mu \nu}$
cannot be determined. Hence, at this point, we have to allow ourselves an 
assumption about the structure of the theory of modified gravity.
In section IV, we discuss two possibilities. 
One possibility is to assume that Birkoff's law is respected, 
as is discussed in Ref.~\cite{birkoff}. We will 
show that this assumption can be translated into conditions
on $E_{\mu \nu}$. In this paper, we also consider a different 
possibility. Since the theory contains the DGP model as 
a special case $n=2$, where the conditions on $E_{\mu \nu}$
are known \cite{KR}, the same conditions on 
$E_{\mu \nu}$ as the DGP may be applied for general $n>2$. Then it turns out that 
the resultant theory has a very similar structure for the quasi-static sub-horizon
perturbations to the DGP model. We comment on 
a potential danger of having a ghost in this case as in the DGP model.
Then in section V we discuss how different forms 
of the modified theory of gravity predict different structure formation 
with the identical background expansion histories of the universe. 
Section VI is devoted to conclusions. 

\section{Covariant effective equations for modified Friedmann equation}
We begin with the DGP model where the covariant effective theory is well known. 
In the DGP model, the effective gravitational equations are 
given by \cite{Maeda:2003ar} 
\be
G_{\mu \nu} = \kappa_{(2)}^4 {}^{(2)}\Pi_{\mu \nu} -E_{\mu \nu},
\ee
where $^{(2)}\Pi_{\mu \nu}$ is 
\be
{}^{(2)}\Pi_{\mu \nu} = -\frac{1}{4} \tilde{T}_{\mu \alpha}
\tilde{T}_{\nu}^{\alpha} +\frac{1}{12} \tilde{T}^{\alpha}_{\alpha}
\tilde{T}_{\mu \nu} + \frac{1}{24}\left[3
\tilde{T}_{\alpha \beta} \tilde{T}^{\alpha \beta}-
(\tilde{T}^{\alpha}_{\alpha})^2\right] g_{\mu \nu}\,,
\label{Pidgp}
\ee
\be
\tilde{T}_{\mu \nu} =  \kappa_4^{-2} G_{\mu \nu}-T_{\mu \nu}\,,
\label{tilT}
\ee
and $\kappa^2_{(2)}$ is given by the five-dimensional Newton constant 
$\kappa^2_{(2)}= 8 \pi G_5$. $E_{\mu \nu}$ is a projection of 
the 5-dimensional Weyl tensor and it is traceless. 
The Bianchi identity imposes 
\be
\nabla^{\mu} E_{\mu \nu} = \kappa_{(2)}^4 \nabla^{\mu} {}^{(2)}\Pi_{\mu \nu},
\label{bianchico}
\ee
while the energy momentum tensor satisfies
\be
\nabla^{\mu} T_{\mu \nu}=0.
\ee
In a homogeneous and isotropic background, the energy momentum tensor is written as 
\be
T^{\mu}_{\:\: \nu} =\mbox{diag}(-\rho, P,P,P).
\ee
From the definition of $\tilde{T}_{\mu \nu}$, Eq.~(\ref{tilT}), 
we get 
\be
T^{\mu}_{\:\: \nu} =\mbox{diag}(-\tilde{\rho}, \tilde{P},\tilde{P},\tilde{P}),
\ee
where
\be
\tilde{\rho} =\frac{3}{\kappa_4^{2}} H^2 - \rho,\quad
\tilde{P}= -\frac{1}{\kappa_4^{2}}(2 \dot{H}+3 H^2)-P.
\label{tilrho}
\ee
Then $^{(2)}\Pi_{\mu \nu}$ is calculated as 
\be
{}^{(2)} \Pi^{\mu}_{\:\: \nu} = \frac{1}{12}
\left(
\begin{array}{cc}
-\tilde{\rho}^2  &  0 \\
0  &  (\tilde{\rho}^2 + 2 \tilde{P} \tilde{\rho}) \delta^i_{\:\: j}\\
\end{array}
\right).
\ee
We can show that this expression satisfies $\nabla^{\mu} {}^{(2)}\Pi_{\mu \nu}=0$,
thus we can set $E_{\mu \nu}=0$ consistently in the background. In fact, it is 
known that non-zero $E_{\mu \nu}$ corresponds to the existence of a black hole 
in the bulk and we can consistently set the black hole mass zero as 
a boundary condition in the bulk.
The $(0,0)$ component of the effective equations gives 
\be
H^2 = r_c^2 \left(H^2 - \frac{\kappa_4^2}{3} \rho \right)^2.
\ee
Then we arrive at the Friedmann equation,
\be
\frac{H}{r_c} = H^2 - \frac{\kappa_4^2}{3} \rho,
\ee
where we have chosen the sign so that the solution yields 
a late time acceleration. In the DGP model, this corresponds 
to the choice of the embedding of the brane in 5-dimensional
spacetime. 

Now we extend this covariant equation to general $n$. 
We can construct a covariant expression for $^{(n)}\Pi_{\mu \nu}$
that yields 
\be
{}^{(n)}\Pi^{\mu}_{\:\: \nu} = \frac{1}{4 \times 3^{n-1}}
\left(
\begin{array}{cc}
-\tilde{\rho}^n  &  0 \\
0  &  \Big[(n-1) \tilde{\rho}^n + n \tilde{P}\rho^{n-1} \Big] \delta^i_{\:\: j}\\
\end{array}
\right), \label{Pib}
\ee
in the background (see Appendix).
Again $\nabla^{\mu} {}^{(n)}\Pi_{\mu \nu}=0$ is satisfied and we can 
set $E_{\mu \nu}=0$ consistently.
Then the $(0,0)$ component of the effective equations gives 
\be
H^2 = r_c^{2(n-1)} \left(H^2 - \frac{\kappa_4^2}{3} \rho \right)^n,
\ee
and the Friedmann equation is obtained as 
\be
\left(\frac{H^2}{r_c^{2(n-1)}}\right)^{1/n} = H^2 - \frac{\kappa_4^2}{3} \rho.
\label{modified}
\ee
It is useful to note that using the Friedmann equation, $\tilde{\rho}$ and $\tilde{P}$,
defined in Eq~(\ref{tilrho}) are rewritten as 
\bea
\tilde{\rho} &=& -\frac{3H^2}{\kappa_4^2} (Hr_c)^{-2(n-1)/n}, \quad \nonumber\\
\tilde{P} &=& \frac{3 H^2}{\kappa_4^2} \left(1+ \frac{2\dot{H}}{3n H^2} \right)
(Hr_c)^{-2(n-1)/n}.
\eea

\section{perturbations}
Now we consider the perturbations. In this paper, we concentrate our attention 
on the scalar perturbations which are relevant for structure formation.
We take the perturbed energy-momentum tensor as
\be
T^{\mu}_{\:\: \nu} = 
\left(
\begin{array}{cc}
- \delta \rho  &  a \delta q_{,i} \\
-a^{-1} \delta q^{,i}  &  \delta P \delta^i_{\:\: j} + \delta \pi^{i}_{\:\: j}\\
\end{array}
\right),
\ee
where 
\be
\delta \pi^{i}_j = \delta \pi^{,i}_{,j}
- \frac{1}{3} \delta^{i}_j \delta \pi^{,k}_{,k}.
\ee
Let us begin with the DGP model. 
The perturbed ${}^{(2)} \Pi^{\mu}_{\:\: \nu}$ is calculated as 
\be
\delta {}^{(2)} \Pi^{\mu}_{\:\: \nu} = \frac{1}{12}
\left(
\begin{array}{cc}
- 2 \tilde{\rho} \delta \tilde{\rho}  &  2 a \tilde{\rho} \delta \tilde{q}_{,i} \\
- 2 a^{-1} \rho \delta \tilde{q}^{,i}  &  2 \left\{ (\tilde{\rho}+\tilde{P}) 
\delta \tilde{\rho} + \tilde{\rho} \delta \tilde{P} \right\}
\delta^i_{\:\: j} 
-(\tilde{\rho}+3 \tilde{P}) \delta \tilde{\pi}^{i}_{\:\: j}\\
\end{array}
\right),
\ee
where
\begin{eqnarray}
\delta \tilde{\rho} &=& \delta \rho + \frac{1}{\kappa_4^2} \delta G^0_0, \quad
\delta \tilde{q}_{,i} = \delta q_{,i} - \frac{1}{a \kappa_4^2} \delta G^{0}_i, \\
\delta \tilde{P} &=& \delta P - \frac{1}{\kappa_4^2} \delta G^T, \quad
\delta \tilde{\pi} = \delta \pi - \frac{1} {\kappa_4^2} \delta G^L.
\end{eqnarray}
and we decomposed the $(i,j)$ component of the perturbed Einstein tensor
as 
\begin{equation}
\delta G^{i}_j = \delta G^L \delta^i_{j} + \delta G^{T,i}_{\:\:\:\:,j}
- \frac{1}{3} \delta^{i}_j \delta G^{T ,k}_{\:\:\:\: ,k}. 
\end{equation}
Perturbations of the divergence of $^{(2)}\Pi_{\mu \nu}$ are calculated as 
\be
\delta (\nabla^{\mu} {}^{(2)}\Pi_{\mu 0})=0,
\ee
\be
\delta(\nabla^{\mu} {}^{(2)} \Pi_{\mu i}) =\frac{(\tilde{\rho}+\tilde{P})}{6} 
\left(\delta \tilde{\rho} -3 a H \delta \tilde{q} + k^2 \delta \tilde{\pi} \right)_{,i},
\ee
where the energy-momentum conservation equations
\be
\dot{\delta \rho} + 3 H (\delta \rho + \delta P) 
+ 3 (\rho+P) \dot{\Phi} -a^{-1}k^2 \delta q =0,
\label{econ0}
\end{equation}
\begin{equation}
\dot{\delta q} + 4H \delta q + a^{-1} 
\left [(\rho+P) \Psi + \delta P - \frac{2}{3} k^2 \delta \pi  \right]=0.
\label{econi}
\ee
are used and the Longitudinal gauge metric is adopted,
\begin{equation}
ds_4^2= -(1+2 \Psi)dt^2 + a^2 (1+2 \Phi) \delta_{ij}dx^i dx^j.
\end{equation}
It is clear that the Bianchi identity Eq.~(\ref{bianchico}) 
cannot be satisfied without perturbations of $E_{\mu \nu}$. 
We can parameterize the scalar perturbations of $E_{\mu \nu}$
as an effective fluid, 
\begin{equation}
\delta E^{\mu}_{\nu} = 
\left(
\begin{array}{cc}
 -\delta \rho_E & a \delta q_{E,i} \\
 -a^{-1} \delta q_E^{,i}  & \delta P_E \: 
 \delta_{ij} + \delta \pi^{i}_{Ej} \\
\end{array}
\right).
\end{equation}
Then the Bianchi identity yields constraint equations for $\delta E_{\mu \nu}$ as
\begin{equation}
\dot{\delta \rho}_E + 3 H (\delta \rho_E+ \delta P_E) - a^{-1} k^2 \delta q_E =0,
\end{equation}
\begin{equation}
\dot{\delta q}_{E} + 4 H \delta q_E + a^{-1} 
\left(\delta P_E - \frac{2}{3} k^2 \delta \pi_E \right)
=-\frac{2}{3} r_c \frac{\dot{H}}{H} 
\left\{
\delta \tilde{\rho} -3 H a \delta \tilde{q} +k^2 \delta \tilde{\pi}
 \right\},
\end{equation}
where background equations are used to rewrite $\tilde{\rho}$ and $\tilde{P}$. 

For general $n$, the covariant expression $^{(n)}\Pi_{\mu \nu}$ yields
\begin{eqnarray}
\delta {}^{(n)} \Pi^{\mu}_{\:\: \nu} = \frac{1}{4 \times 3^{n-1}}
\left(
\begin{array}{cc}
- n \tilde{\rho}^{n-1} \delta \tilde{\rho}  &  n a \tilde{\rho}^{n-1} 
\delta \tilde{q}_{,i} \\
- n a^{-1} \tilde{\rho}^{n-1} \delta \tilde{q}^{,i}  &  n \left\{(n-1) 
\tilde{\rho}^{n-2} (\tilde{\rho}+\tilde{P}) \delta \tilde{\rho} 
+ \rho^{n-1} \delta \tilde{P} \right\}
\delta^i_{\:\: j} 
+f_n(\tilde{\rho},\tilde{P}) \delta \pi^{i}_{\:\: j}\\
\end{array}
\right),
\end{eqnarray}
where $f_n(\tilde{\rho},\tilde{P})$ is not determined by the requirement that 
the background $^{(n)}\Pi_{\mu \nu}$ is given by Eq.~(\ref{Pib}). 
Again, the $0$-component of the divergence of ${}^{(n)} \Pi_{\mu \nu}$
satisfies
\be
\delta (\nabla^{\mu} {}^{(n)}\Pi_{\mu 0})=0,
\ee
automatically, but the $i$-component is non-zero. Thus we have to 
introduce perturbations of $E_{\mu \nu}$ and the Bianchi 
identity gives constraint equations for $E_{\mu \nu}$. 
In this paper, we will demand that $f_n(\tilde{\rho}, \tilde{P})$ is of the form 
$f_n(\tilde{\rho}, \tilde{P}) = b_1 \tilde{\rho}^{n-1} + b_2 \tilde{\rho}^{n-2} \tilde{P}$ 
like the other components. This requirement fixes $f(\tilde{\rho}, \tilde{P})$ as 
\be
f_n(\tilde{\rho}, \tilde{P})= n \left(-\frac{3}{2}n + \frac{5}{2} \right) \tilde{\rho}^{n-1}
- \frac{3}{2} n(n-1) \tilde{\rho}^{n-2} \tilde{P}.
\label{Piani}
\ee
We have explicitly checked this formula for $n=3,4$ and $5$ (see Appendix). 

Now we can write down the perturbed Einstein equations;
\begin{equation} 
\left[1- \frac{1}{n \alpha(Hr_c)} \right] \delta G^{0}_0 = - \kappa_4^2 \delta \rho
+ \frac{\kappa_4^2}{n \alpha(Hr_c)} \delta \rho_{E},
\end{equation}
\begin{equation}
\left[1- \frac{1}{n \alpha(Hr_c)} \right] \delta G^{0}_i = \kappa_4^2 
a \delta q_{,i} - \frac{\kappa_4^2}{n \alpha(Hr_c)} a \delta q_{E,i},
\end{equation}
\begin{equation}
 \left[   
1- \frac{1}{n \alpha(Hr_c) \left(1 + 
\frac{n-1}{n} \frac{\dot{H}}{H} \right)} \right] \delta G^{T}
= \kappa_4^2 \delta \pi - 
\frac{\kappa_4^2 \delta \pi_{E}}{ n \alpha(Hr_c) 
\left(1+ \frac{n-1}{n} \frac{\dot{H}}{H}\right)},
\end{equation}
where we defined 
\be
\alpha(Hr_c)=(Hr_c)^{2(n-1)/n}.
\ee
The constraint equations for $\delta E_{\mu \nu}$ are given by
\begin{equation}
\dot{\delta \rho}_E + 4 H \delta \rho_E - a^{-1} k^2 \delta q_E =0,
\end{equation}
\begin{equation}
\dot{\delta q}_{E} + 4 H \delta q_E + a^{-1} 
\left(\delta P_E - \frac{2}{3} k^2 \delta \pi_E \right)
=-\frac{2}{3}\frac{\dot{H}}{H^2} \alpha(Hr_c) (n-1)a^{-1}  F,
\end{equation}
\be
F=  \frac{\rho \bigtriangleup}{1 - n \alpha(Hr_c)} 
+\frac{\delta \rho_E-3Ha\delta q_E}{n \alpha(Hr_c)-1} + 
\frac{k^2 \delta \pi_E}{n \alpha(Hr_c) \left(1+\frac{n-1}{n} \frac{\dot{H}}{H}\right)-1},
\ee
where
\be
\rho \bigtriangleup = \delta \rho -3 H a \delta q.
\ee
Together with the energy-momentum conservation Eqs.~(\ref{econ0}) and 
(\ref{econi}), these equations are the basis for the analysis of perturbations 
around the modified background Eq.~(\ref{modified}). 

\section{Behaviour of perturbations}
In this section, we study the behavior of the quasi-static perturbations on 
subhorizon scales. With this approximation, the constraint equations 
for $E_{\mu \nu}$ become 
\be
\delta q_E =0, \quad \delta P_E - \frac{2}{3} k^2 \delta \pi_E 
=-\frac{2}{3}\frac{\dot{H}}{H^2} \alpha(Hr_c)(n-1) a^{-1}  F.
\label{constraint-E}
\ee
The key equations obtained from the perturbed Einstein equations are 
\begin{eqnarray}
&&\frac{k^2}{a^2} \Phi = \frac{\kappa_4^2}{2} \left( \frac{n \alpha(Hr_c)}{n\alpha(Hr_c) -1}
\right) \left (\rho \bigtriangleup - \frac{\delta \rho_{ E}}{n\alpha(Hr_c)} \right), \label{einstein-t}\\
&&\Phi+\Psi = \kappa_4^2 \frac{1}{n \alpha(Hr_c) \left(1+\frac{n-1}{n} \frac{\dot{H}}{H^2}\right)-1} 
\, a^2 \delta
\pi_{ E}\,. \label{einstein-ijt}
\end{eqnarray}
From energy-momentum conservation, the evolution equation
for matter over-density is obtained as 
\be
\ddot{\bigtriangleup}+ 2H \dot{\bigtriangleup}=-\frac{k^2}{a^2} \Psi.
\label{dele}
\ee
The effective equations are not closed and we need to assume 
equations of state for $E_{\mu \nu}$. Thus, at this point, we have to 
assume a structure of the theory of modified gravity. Within our approximations, 
we need two equations of state, such as 
\be
\delta P_E = W(Hr_c) \delta \rho_E, \quad  \delta \rho_E = C(Hr_c) k^2 \delta \pi_E,
\ee
to close the equations. In the following, we will consider two possibilities. 

\subsection{Birkoff's law}
In the background, we set $E_{\mu \nu}=0$, thus it might be natural to 
consider conditions given by
\be
\delta \rho_E =\delta P_E =0. 
\ee
We should stress that it is impossible to take $\delta \pi_E=0$ due to the 
constraint Eq.~(\ref{constraint-E}). 
It is also important to mention that, in the DGP model, these conditions imply the 
divergent behaviour of perturbations in the bulk and so are unphyical. 
With these conditions, we can solve the constraint equations Eq.~(\ref{constraint-E})
to get 
\be 
k^2 \delta \pi_E =-(n-1) \left(\frac{\dot{H}}{H^2} \alpha(Hr_c) \right)
\frac{n \alpha(Hr_c) \left( 1+ \frac{n-1}{n} \frac{\dot{H}}{H^2} \right)-1}
{\left[n\alpha(Hr_c) -1 \right]^2} \rho \bigtriangleup.
\ee
Then from Eqs.~(\ref{einstein-t}) and (\ref{einstein-ijt}), 
the solutions for metric perturbations are obtained as 
\bea
\frac{k^2}{a^2} \Phi &=& \frac{\kappa_4^2}{2} \frac{n \alpha(Hr_c)}
{n \alpha(Hr_c)-1} \rho \bigtriangleup, \\
\frac{k^2}{a^2} \Psi &=& \kappa_4^2 \left\{ -\frac{1}{2} \frac{n \alpha(Hr_c)}
{n \alpha(Hr_c)-1} -(n-1) \frac{\dot{H}}{H^2} \frac{\alpha(Hr_c)}
{(n\alpha(Hr_c)-1)^2} \right\} \rho \bigtriangleup.
\eea
Using the background equations for a dust dominated universe,
\be
\dot{H}= \frac{3n H^2}{2} \frac{1-\alpha(Hr_c)}{n\alpha(Hr_c)-1},
\ee
the above solutions can be written as 
\bea
\frac{k^2}{a^2} \Phi 
&=& \frac{\kappa_4^2}{2} g'(x) \rho \bigtriangleup, \\
\frac{k^2}{a^2} \Psi 
&=& -\frac{\kappa_4^2}{2} \Big[ g'(x) +3 x g''(x) \Big] \rho \bigtriangleup,
\eea
where $g(x)$ is defined from the modified Friedmann equation Eq.~(\ref{modified})
\be
H^2 = H_0^2 g(x), \quad x = \frac{\kappa_4^2 \rho}{3 H_0^2}.
\ee
Then the evolution equation for over-density is given by 
\be
\ddot{\bigtriangleup}+2 H \dot{\bigtriangleup} 
= \frac{\kappa_4^2}{2} \Big[ g'(x)+3 x g''(x) \Big] \rho \bigtriangleup.
\label{evoldelb}
\ee
These solutions are the same as the results obtained in Ref.~\cite{birkoff}
where Birkoff's law is assumed to be respected. If Birkoff's
law is respected, the dynamics of a spherical symmetric collapsing dust 
shell with radius $r=R(t)$ can be derived from the 
Friedmann equation. The time derivative of the Friedmann equation yields
\be
\frac{\ddot{a}}{a} = H_0^2 
\left[ g(x)-\frac{3}{2}x g'(x) \right],
\ee
which is valid also for non-flat 3-space. Then Birkoff's
law implies that the dynamics of $R(t)$ is given by
\be
\frac{\ddot{R}(t)}{R(t)} = H_0^2 
\left[ g(x)-\frac{3}{2}x g'(x) \right],
\ee
where
\be
x=\frac{\kappa_4^2 \rho_{shell}}{3 H_0^2}, \quad 
\rho_{shell} =\frac{3 M}{4 \pi R(t)^3}, 
\ee
and $M$ is the total mass contained in the shell. 
The over-density is defined by
\be
\Delta = \frac{\rho_{shell} - \rho}{\rho}.
\ee
Initially, the shell is expanding just due to the expansion 
of the universe, so at an initial time  $R(t) =a(t) R_0$ 
and $\rho_{shell}=\rho$. The conservation of mass in the shell means 
that $R(t)$ and $\bigtriangleup$ are related from the condition 
$R^3(t) \rho_{shell}= a^3 R_0^3 \rho$ as
\be
R(t)=a(t) R_0 \Big[ 1+ \bigtriangleup(t) \Big]^{-1/3}.
\ee
Then the equation for $R(t)$ can be rewritten into the equation 
for $\bigtriangleup$. By linearizing the equation, we arrive at 
Eq.~(\ref{evoldelb}).

\subsection{DGP-like model}
In the DGP model, correct conditions for $E_{\mu \nu}$ are
obtained by solving the 5D perturbations and imposing the 
regularity condition in the bulk \cite{KR}. 
The conditions are obtained as
\be
\delta P_E = \frac{1}{3} \delta \rho_E, \quad 
\delta \rho_E= 2 k^2 \delta \pi_E,
\ee
where the former condition comes from the fact that 
$E_{\mu \nu}$ is a projection of 5D Weyl tensor 
and it is traceless. Let us apply these conditions 
for general $n$. Solving the constraint equations Eq.~(\ref{constraint-E}),
we get
\be
\delta \rho_E = \frac{2 \left[ n\alpha(Hr_c) \left(   
1+ \frac{n-1}{n} \frac{\dot{H}}{H^2} \right)-1  \right]}
{3 \left[ n\alpha(Hr_c) \left(   
1+ \frac{2(n-1)}{3n} \frac{\dot{H}}{H^2}  \right)-1 \right]} \rho \bigtriangleup.
\ee
Then the solutions for metric perturbations are obtained as 
\begin{eqnarray}
\frac{k^2}{a^2} \Phi &=& \frac{\kappa_4^2}{2} \left(1- \frac{1}{3 \beta} \right)
\rho \triangle \label{solphi}\\
\frac{k^2}{a^2} \Psi &=& - \frac{\kappa_4^2}{2} 
\left(1 + \frac{1}{3 \beta} \right) \rho \triangle \label{solpsi}
\end{eqnarray}
where
\begin{equation}
\beta =1 -n \alpha(Hr_c) \left(1+ \frac{2(n-1)\dot{H}}{3n H^2} \right).
\end{equation}
The evolution equation for the linear over-density is given by
\begin{equation}
\ddot{\bigtriangleup} + 2 H \dot{\bigtriangleup}=\frac{\kappa_4^2}{2} \left(1 +
\frac{1}{3 \beta} \right) \rho \bigtriangleup\,.
\label{evoldeld}
\end{equation}
For $n=2$, these results reproduce those obtained in Ref.~\cite{Lue:2004rj}.

Linearized perturbations are described by a scalar tensor theory 
with the Brans-Dicke (BD) coupling given by 
\be
\omega=\frac{3}{2}(\beta-1),
\ee
where the scalar field originates from the scalar polarization of the 
graviton.
Unfortunately, this shows a potentially serious problem of the model. 
In the Einstein frame where kinetic terms for the spin-2 graviton and 
the scalar mode are diagonalized, 
the scalar mode has a wrong sign for its kinetic term if 
\be
\omega+\frac{3}{2} = \frac{3}{2} \beta<0.
\ee
We can show that $\beta$ is always negative in a 
phyical situation, thus this indicates that we have a ghost-like excitation.
This can be confirmed more rigorously in the DGP model if the 
brane is purely de Sitter spacetime. In this case, the condition
for $\beta <0$ implies $1/2 < Hr_c$ and this is precisely the 
condition to have a ghost derived in Ref.~\cite{problems2}. 
We will come back to this problem in the conclusion.

\section{Structure formation}
In general, the linear growth factor is determined by Eqs.~(\ref{einstein-t}),
(\ref{einstein-ijt}) and (\ref{dele}). 
It is manifest that the growth factor crucially depends on $\delta \rho_E$ and 
$\delta \pi_E$. Hence the structure of the modified gravity, which
is described by the equations of state of $E_{\mu \nu}$, is encoded 
in the linear growth factor. In order to have an insight into how the 
different structures of the modified gravity change the linear 
growth factor, we show the linear growth factor for the Birkoff's
law models and the DGP-like models studied in the previous section (Fig.1). 
We also compare them with the growth factor in dark energy models with 
identical expansion histories. As seen from Fig.2, the growth factor 
is different from the one in dark energy models with identical 
expansion histories. This fact can be used to distinguish between
modified gravity models and dark energy models \cite{Lue:2004rj}, \cite{birkoff},
\cite{Linder}, \cite{wrong}. 

\begin{figure}[t]
\centerline{
\includegraphics[width=16cm]{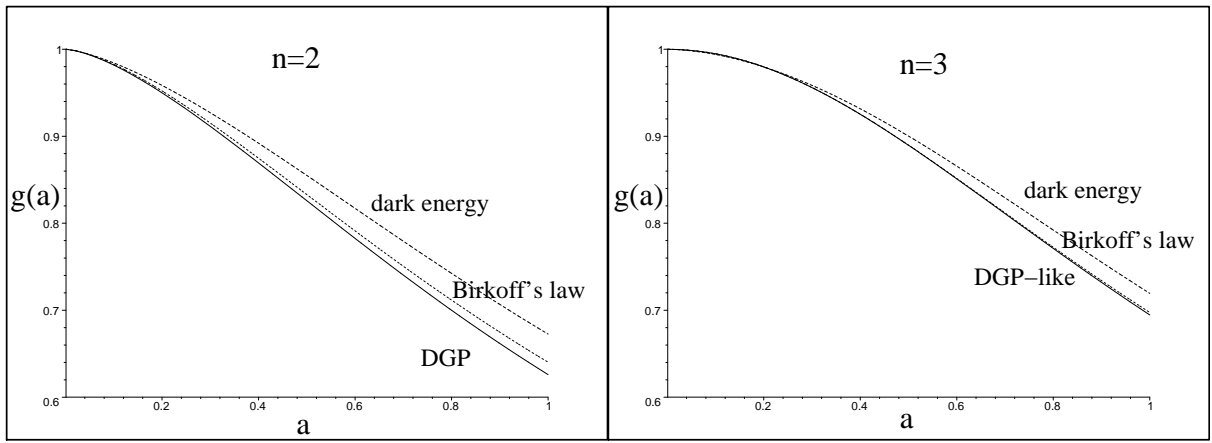}}
\caption{The growth history $g(a)=\bigtriangleup(a)/a$ is shown for 
$n=2$ (left panel) and $n=3$ (right panel). From top to bottom, 
the lines show the growth history of dark energy models with 
identical expansion history, the Birkoff's law model and the 
DGP model. We set the density parameter for matter today as 
$\Omega_{m0}=0.3$.}
\label{fig:fig1}
\end{figure}

\begin{figure}[t]
\centerline{
\includegraphics[width=16cm]{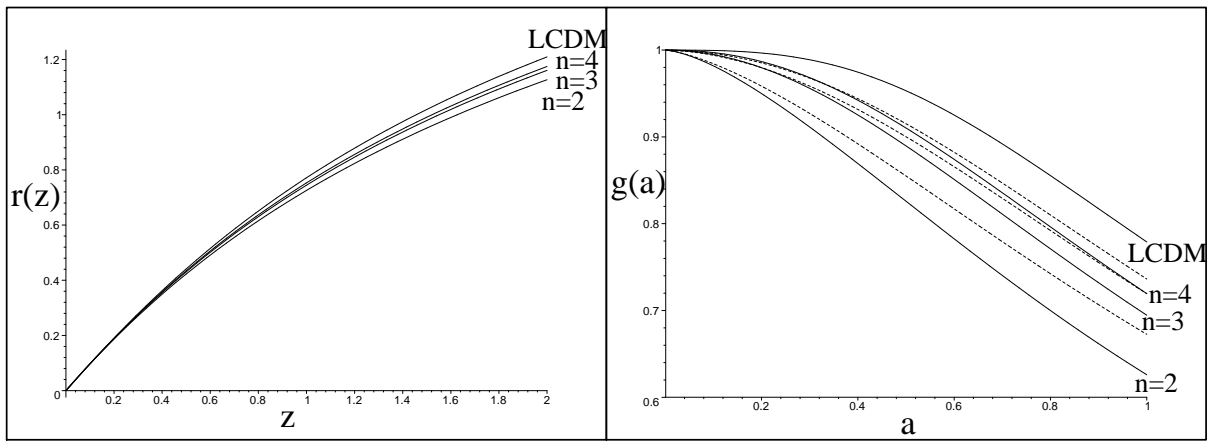}}
\caption{In the left panel, the expansion histories $r(z)=\int^z_0 dz H(z)^{-1}$ 
are shown for LCDM, $n=4$, $n=3$ and $n=2$ from top to bottom. 
In the right panel, the growth factor $g(a)=\bigtriangleup(a)/a$ is shown. 
The solid lines are LCDM, DGP-like models for $n=4$, $n=3$ and $n=2$ from top to 
bottom and the dotted lines show corresponding dark energy models with the identical 
expansion histories}
\label{fig:fig2}
\end{figure}

The linear growth factor is the basis for
tests of modified gravity against structure formation observations. But one
also needs the metric perturbations \cite{KR}. In the Birkoff's law models, 
\bea
\frac{k^2}{a^2}(\Phi+\Psi) &=& -\kappa_4^2 (n-1) 
\frac{\dot{H}}{H^2} \frac{ \alpha(Hr_c)}{[n \alpha(Hr_c)-1]^2} \rho \bigtriangleup, \\
\label{iswb}
\frac{k^2}{a^2}(\Phi-\Psi)  &=& 
\kappa_4^2 \left[ \frac{n \alpha(Hr_c)}{n \alpha(Hr_c)-1} + (n-1) \frac{\dot{H}}{H^2}
\frac{\alpha(Hr_c)}{(n \alpha(Hr_c)-1 )^2} \right]
\rho \bigtriangleup.
\eea
On the other hand, in the DGP-like models,
\bea
\frac{k^2}{a^2}(\Phi +\Psi) &=& -\kappa_4^2 \frac{1}{3 \beta} \rho \bigtriangleup, \\
\label{iswd}
\frac{k^2}{a^2}(\Phi-\Psi)  &=& \kappa_4^2 \rho \bigtriangleup.
\eea
Equations~(\ref{iswb}) and (\ref{iswd}) basically determine the
integrated Sachs-Wolfe (ISW) and weak lensing effects. 
In the Birkoff's law model, this equation receives an additional 
correction from the modified gravity compared with general relativity \cite{birkoff}. 
Thus the growth factor 
is not enough to predict the ISW effects and weak lensing effects. 
Fig.~3 shows $k^2(\Phi-\Psi)/a^2 \kappa_4^2 \rho \bigtriangleup$. 
Although for $n>2$, the growth factor is almost identical 
in both models, the ISW effects and weak lensing effects are different due to the time 
variation of the functions in Eq.~(\ref{iswb}) 
in the Birkoff's law model. 
This manifests the need for full solutions for the 
metric perturbations in order to predict structure formation
in modified gravity models.

\begin{figure}[t]
\centerline{
\includegraphics[width=8cm]{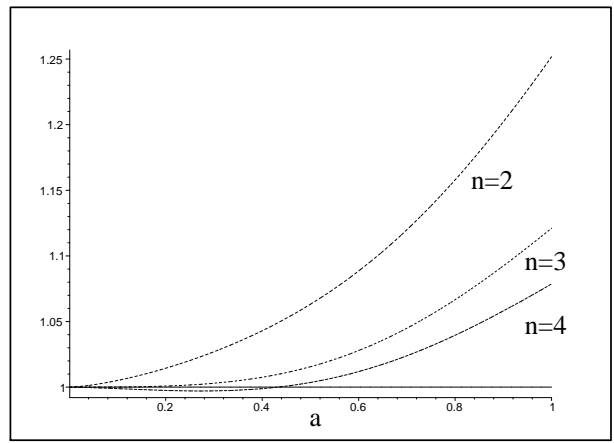}}
\caption{$k^2(\Phi-\Psi)/a^2 \kappa_4^2 \rho \bigtriangleup$ is shown 
for $n=2$, $n=3$, $n=4$ in the Birkoff's law models. 
$k^2(\Phi-\Psi)/a^2 \kappa_4^2 \rho \bigtriangleup=1$ in 
the DGP-like models and dark energy models in general relativity.}
\label{fig:fig3}
\end{figure}

\section{conclusion}
In this paper, we studied structure formation in phenomenological 
models in which the modified Friedmann equaiton is given by Eq.~(\ref{modified}). 
In order to study the perturbations around this modified background, 
we constructed simple covariant equations which reproduce the 
modified Friedmann equation in the cosmological background. 
By requiring that the Bianchi identity is satisfied, a possible form of the effective 
equations is restricted. At a perturbative level, we have to introduce an unknown 
term $E_{\mu \nu}$ in the effective equations to satisfy the Bianchi identity. 
The Bianchi identity imposes the constraint equations for $E_{\mu \nu}$, 
but they are not closed. Thus the structure of the modified theory of gravity 
is encoded in the additional conditions on $E_{\mu \nu}$, which are needed to 
close the effective equations. 

In this paper, we analyze the quasi-static sub-horizon perturbations 
relevant for structure formation. 
We consider two possibilities. One is a model in which 
Birkoff's law is respected and the other is a model in which 
the perturbations can be described by scalar-tensor gravity as in the 
DGP model. These models are realized by particular choices of the 
equation of state for $E_{\mu \nu}$. We demonstrate how the growth factor 
depends on the assumed structure of the modified gravity. Interestingly, 
in models with $n>2$, the two models give almost identical growth factor. 
However, this does not imply the observational consequences are the same. 
The growth factor is determined by the solutions for $\Psi$, but the 
ISW and weak lensing effects are determined by the combination $\Phi-\Psi$. 
We find that the two models predict different behaviours of $\Phi - \Psi$. 
This manifests the need for knowledge of metric perturbations in order 
to address structure formation in modified gravity models $-$ the 
parameterization of the growth factor is not enough. 

In this paper, we only consider linear perturbations. It is crucial 
to study non-linear perturbations since the solar system constraints
require that the theory must be close to general relativity at solar system
scales. In the Birkoff's law model, it is shown that below the scale 
$r_* =(r_g/H_0^2)^{1/3}$, the deviation from general relativity is small
and the solar system constraints can be evaded \cite{birkoff}. 
This is also true in the DGP model \cite{nonlinear}. Thus we suspect the 
non-linear recovery of general relativity is quite generic in these phenomenological 
models. Detailed features of the transition from modified gravity to general 
relativity depend on the structure of the modified theory of gravity. If we can 
extract an information on the transition from linear theory to non-linear theory from 
observations such as weak lensing and cluster mass function, this also provides  
an interesting probe of the structure of the modified gravity.
The effective equations are valid for non-linear physics, but we have to generalize 
the conditions on $E_{\mu \nu}$ to non-linear physics. It would be interesting 
to study the non-linear dynamics such as a spherical collapse based on 
the effective equations and we will leave this issue as a future work.

The obvious outstanding issue is to find a fundamental theory that 
reproduces the effective equations in this paper. Only 
the underlying theoretical model is capable of fixing unambiguously 
the conditions for $E_{\mu \nu}$ and hence the growth rate . 
Moreover, in a self-accelerating universe in the DGP model, the 
additional suppression of the growth rate compared with a dark 
energy model is 
tightly connected with the pathology of the model. In this model, 
the scalar mode of the graviton behaves like a ghost. This is the origin 
of the additional suppression of the growth rate, because the ghost 
gives a repulsive force and this prevents the CDM over-density 
from collapsing. Our analysis may indicate a possibility of 
avoiding the ghost by changing the boundary condition for $E_{\mu \nu}$, 
because the conclusion of having a theory with a negative BD parameter
comes from a specific condition on $E_{\mu \nu}$ based on 
regularity in the bulk. This modification may be achieved by 
an introduction of a second brane in the bulk \cite{tanaka}. 
The growth rate is sensitive to this modification. Hence it is 
crucial to develop a consistent theory for the modified gravity 
and, once we have a consistent theory, structure formation measures 
become essential to test the theory against observations. 

\[ \]
{\bf Acknowledgments:}
We would like to thank R.~Maartens for discussions and 
a careful reading of this manuscript. 
KK is supported by PPARC.

\appendix
\section{Construction of covariant tensor $^{(n)}\Pi_{\mu \nu}$}
In this appendix, we construct an expressions for $^{(n)}\Pi_{\mu \nu}$. 
We first construct $^{(n)}\Pi_{\mu \nu}$ as a function of $T_{\mu \nu}$
and then replace $T_{\mu \nu}$ by $\tilde{T}_{\mu \nu}$. 
We start from the simplest case where $^{(n)}\Pi_{\mu \nu}$ is 
a quadratic function of the energy-momentum tensor $T_{\mu \nu}$. 
In this case the answer is known from the effective equations in the 
DGP model, i.e. Eq.~(\ref{Pidgp}), but it is instructive to reconstruct 
this expression. 
The general form of $^{(2)}\Pi_{\mu \nu}$ is given by
\be
{}^{(2)}\Pi^{\mu}_{\:\: \nu} =A_1 T_{\mu \alpha} T^{\alpha}_{\:\:\nu} +A_2 T T^{\mu}_{\:\: \nu}
+A_3 T_{\alpha \beta}T^{\alpha \beta} \delta^{\mu}_{\:\: \nu} + A_4 T^2  \delta^{\mu}_{\:\: \nu}, 
\ee
where $T=T^{\alpha}_{\:\: \alpha}$.
In the homogeneous and isotropic universe, 
we require that ${}^{(2)}\Pi^{0}_{\:\: 0} \propto \rho^2$ and $^{(2)}\Pi_{\mu \nu}$ satisfies
\be
\nabla^{\nu} {}^{(2)}\Pi_{\mu \nu}=0.
\ee
Then $^{(2)}\Pi_{\mu \nu}$ must be of the form
\be
{}^{(2)} \Pi^{\mu}_{\:\: \nu} =
\left(
\begin{array}{cc}
B_1 \rho^2  &  0 \\
0  &  (B_2 \rho^2 + B_3 P \rho) \delta^i_{\:\: j}\\
\end{array}
\right).
\label{form}
\ee
This requirement is sufficient to determine the coefficients $A_1,A_2,A_3$ and $A_4$ 
up to an overall normalization;
\be
A_1=6 A_4,\quad A_2 =-2 A_4, \quad A_3=-3 A_4. 
\ee
Then $^{(2)}\Pi_{\mu \nu}$ is given by
\be
{}^{(2)} \Pi^{\mu}_{\:\: \nu} = C_2
\left(
\begin{array}{cc}
-\rho^2  &  0 \\
0  &  (\rho^2 + 2 P \rho) \delta^i_{\:\: j}\\
\end{array}
\right),
\ee
where $C_2=-2 A_4$.
Interestingly, the Bianchi identity $\nabla^{\mu} {}^{(2)}\Pi_{\mu 0}=0$ is 
automatically satisfied. 

For $n=3$, the general form of ${}^{(3)} \Pi_{\mu \nu}$ is given by
\be
{}^{(3)}\Pi^{\mu}_{\nu} = A_1 T^{\mu}_{\alpha} T^{\alpha}_{\beta} T^{\beta}_{\nu}
+A_2 T^{\alpha}_{\beta} T^{\beta}_{\alpha} T^{\mu}_{\nu} 
+A_3 T T^{\mu}_{\alpha} T^{\alpha}_{\nu} + A_4 T^2 T^{\mu}_{\nu}
+A_5 T^{\alpha}_{\beta} T^{\beta}_{\gamma} T^{\gamma}_{\alpha} \delta^{\mu}_{\nu}
+A_6 T T^{\alpha}_{\beta} T^{\beta}_{\alpha} \delta^{\mu}_{\nu}
+A_7 T^3 \delta^{\mu}_{\nu}.
\ee
The same requirement as Eq.~(\ref{form}) fixes the parameters as 
\be
A_1=6A_2-54A_7,\;\;A_3=9 A_7,\;\;A_4=-A_2+3A_7,\;\;A_5=-6A_2+45A_7,\;\;
A_6=2A_2-18A_7.
\ee
Then ${}^{(3)} \Pi^{\mu}_{\nu}$ is determined by
\be
\Pi^{\mu}_{\:\: \nu} = C
\left(
\begin{array}{cc}
-\rho^3  &  0 \\
0  &  (2\rho^3 + 3 P \rho^2) \delta^i_{\:\: j}\\
\end{array}
\right),
\ee
where $C=2 A_2 -14 A_7$. Again the Bianchi identity is automatically 
satisfied. Now let us consider the perturbations.
Non-trivial components are calculated as 
\bea
\delta {}^{(3)} \Pi^{0}_{\:\: i} &=& (6 A_2 -42 A_7) \rho^2 \delta q_{,i}, \nonumber\\
\delta {}^{(3)} \Pi^{i}_{TT \;j} &=& \Big[
(12 A_2 -81 A_7) P^2 + (6A_2 -36A_7) \rho P + 3 A_7 \rho^2 \Big] \delta \pi^{i}_{\:\: j},
\eea
where $\delta {}^{(3)}\Pi^{i}_{TT\;j}$ is the transverse-traceless part of 
the $(i,j)$ component.
In the main text, we demanded that 
\be
12 A_2 - 81 A_7 =0.
\ee
This completely fixes all components except for an overall normalization
and we can verify the formula Eq~(\ref{Piani}). 

We can continue the same procedure for general $n$. We have checked the case 
for $n=4$ and $n=5$. We do not explicitly show the results because the 
formula is very lengthy. For $n=4$, there are $12$ coefficients in the 
general form and for $n=5$, there are $19$ coefficients.
In all cases, a requirement similar to (\ref{form}) is sufficient to 
determine the form of $^{(n)}\Pi_{\mu \nu}$ as  
\be
{}^{(n)}\Pi^{\mu}_{\:\: \nu} = C_n
\left(
\begin{array}{cc}
-\rho^n  &  0 \\
0  &  \Big[ (n-1)\rho^n + n P \rho^{n-1} \Big] \delta^i_{\:\: j}\\
\end{array}
\right).
\ee
The Bianchi identity $\nabla^{\mu} {}^{(n)} \Pi_{\mu 0}=0$ is automatically satisfied.
Then we can compute the perturbations and get  
\bea
\delta {}^{(n)}\Pi^{0}_{\:\: i} &=& n C_n \rho^{n-1} \delta q_{,i}, \nonumber\\
\delta {}^{(n)}\Pi^{i}_{TT\:j} &=& f_n(\rho, P)\delta \pi^{i}_{\:\:j}.
\eea
In this paper, we impose the constraint that 
\be
f_n(\rho, P) = b_1 \rho^{n-1} + b_2 \rho^{n-2} P.
\ee
This gives $n-1$ conditions. We checked that, for $n=4$ and $n=5$, these conditions 
can be imposed consistently and we get 
\be
f_4(\rho, P)= -C_4 (14 \rho^3 +18 \rho P), \quad 
f_5(\rho ,P)= -C_5 (25 \rho^4 + 30 \rho^3 P).
\ee
These results can be summarized as 
\be
f_n(\rho, P)= C_n \left[ n \left(-\frac{3}{2}n + \frac{5}{2} \right) \rho^{n-1}
- \frac{3}{2} n(n-1) \rho^{n-2} P \right].
\ee

\end{document}